# Variable zoom digital in-line holographic microscopy

**Martín Sanz**

*Departamento de Óptica y de Optometría y Ciencias de la Visión. Facultad de Física.*
*Universidad de Valencia. C/ Doctor Moliner 50, 46100 Burjassot, Spain*

**Maciej Trusiak**

*Warsaw University of Technology, Institute of Micromechanics and Photonics, 8 Sw. A.*
*Boboli St., 02-525 Warsaw, Poland*

**Javier García**

*Departamento de Óptica y de Optometría y Ciencias de la Visión. Facultad de Física.*
*Universidad de Valencia. C/ Doctor Moliner 50, 46100 Burjassot, Spain*

**Vicente Micó\***

*Departamento de Óptica y de Optometría y Ciencias de la Visión. Facultad de Física.*
*Universidad de Valencia. C/ Doctor Moliner 50, 46100 Burjassot, Spain*

\* Corresponding author: Vicente Mico, vicente.mico@uv.es

# Abstract

We report on a novel layout providing variable zoom in digital in-line holographic microscopy (VZ-DIHM). The implementation is in virtue of an electrically tunable lens (ETL) which enables to slightly shift the illumination source axial position without mechanical movement of any system component. Magnifications ranging from ~15X to ~35X are easily achievable using the same layout and resulting in a substantial variation of the total field of view (FOV). The performance of the proposed setup is, first, validated using a resolution test target where the main parameters are analyzed (theoretically and experimentally) and, second, corroborated analyzing biological sample (prostate cancer cells) showing its application to biomedical imaging.

**Keywords:** Phase retrieval, Digital image processing, Lensless microscopy, Coherence imaging, electrically tunable lenses.

# 1. Introduction

Lensless microscopy is emerging nowadays as a vividly blossoming technology with growing interest and strong potential in biology/biomedicine where imaging is performed using a lensfree configuration [1-4]. This type of lensless digital holographic microscopy (LDHM) derives from a digital implementation of the Gabor's seminal invention [5, 6] where a point source of coherent light illuminates the sample and the in-line diffracted wavefront is recorded by a digital sensor [7].

Essentially, LDHM can be implemented following two opposite layouts [1], both of them offering different capabilities and applications. One implementation places the sample in close proximity with the illumination point source and the digital sensor farther in comparison [7-14], while the opposite happens in the other [15-19]. The former is commonly known as digital in-line holographic microscopy (DIHM) and introduces a magnification factor (ranging typically from 5X to 20X) by considering the geometrical projection of the sample's diffraction pattern at the digital sensor plane. Although higher values has been reported [20-21], it commonly provides similar properties as in classical DHM regarding field of view (FOV) and resolution limit when considering objectives up to medium numerical aperture (NA) values (~0.4-0.5 NA). The latter is known as on-chip microscopy and it defines no geometric magnification (~1X range) but the allowable FOV becomes extremely huge (~25 mm$^2$) since the whole detector sensitive area is available. However, it provides a modest resolution (incoming from ~0.2 NA range) imposed mainly by the sampling limit (pixel size) of the digital sensor.

Evolutions (in the sense of completeness) of LDHM were reported in recent years. For instance, the two opposite configurations have been merged together into a single one allowing dual and simultaneous lensless imaging mode without altering the intrinsic characteristics and advantages of each one of them [22]. The combination was possible due to the wavelength multiplexing at the illumination/detection stages: green light was used to provide a wide FOV image coming from a 1X-range magnification factor and blue/red illuminations provided a magnified image of the object with higher resolution after applying MISHELF microscopy algorithm [13, 14]. And magnifications lower than 1 were achieved in LDHM in an effort to match the object size with the detector area [23, 24]. Thus, objects having larger sizes than the detector area can be fully imaged without adding any optical components in the layout (such as zooming lenses) by simply changing the illumination vergence (from divergent to convergent) thus preserving the setup simplicity.

In this manuscript, we present a novel implementation in digital in-line holographic microscopy (DIHM) where an electrically tunable lens (ETL) is introduced to modify the axial position of the illumination source. This shift directly impacts in the layout magnification as well as in the available FOV and a continuous variation of both parameters can be easily achieved without neither replacement nor mechanical movement of any component in the optical layout. Strong variation in the magnification factor (from ~15X to ~35X, approximately) and FOV (a factor of around 5) are produced by electrically switching on/off the tunable lens utilizing its full voltage range. The ETL is used for changing the magnification thus optical imaging is still performed without lenses in a lens-free setup.

Nonetheless, ETLs have been applied in microscopy during the last decade, e.g., to improve axial scanning speed in acousto-optical deflection-based two-photon microscopy [25] or in 3D light-sheet microscopy [26]. They were also employed to provide high-speed transport-of-intensity equation quantitative phase microscopy [27], reduce the coherent disturbances in quantitative phase images in self-interference DHM [28], extend the depth of field in optical diffraction tomography [29], enable synthetic aperture super-resolved imaging without assisted reference beam for phase retrieval [30], allow phase distortion compensation in DHM [31], and ensure autofocusing stabilization in total internal reflection microscopy [32], just to cite some examples. But all those approaches yield in classical microscopy setups where objectives are used for imaging.

## 2. System layout description.

Figure 1 presents a scheme of the experimental layout for the proposed variable zoom DIHM (VZ-DIHM). A fiber optic coupled laser diode (Blue Sky Research, SpectraTec 4 STEC4, 405nm) is collimated (CL, QiOptiq achromatic doublet, 120 mm focal length) and directed to a focusing lens (FL, Thorlabs aspheric lens, 0.5 NA) that provides the illumination point source needed in DIHM. The object/sample is placed at z distance (z ≅ 0.5 mm) from the source and the digital sensor (Mightex USB3.0 monochrome camera, board level, 2560 x 1920 pixels, 2.2 μm pixel size) at d distance (d ≅ 12.5 mm) from the sample. Both distances are measured by micrometric stages at the lab. Theoretically and using this geometrical parameters, the layout magnification M can be computed from

$$M = \frac{z+d}{z} \qquad (1)$$

As M = 26X, the imaging system NA equals to NA = sin{arctan[(2.560x2.2)/(2x12.5)]} = 0.22, so the resolution limit results in $\rho = \lambda/NA = 1.85$ μm.

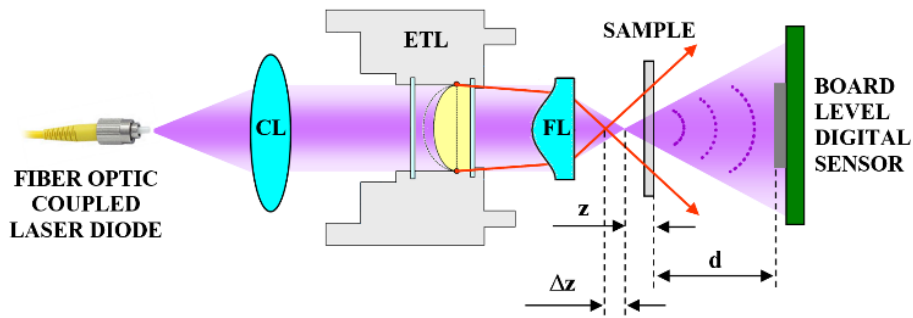

Figure 1. Optical layout for the proposed variable magnification DIHM. CL, condenser lens; ETL, electrically tunable lens; FL, focusing lens.

The ETL (Optotune EL-10-30-C-VIS-LD-MV) is placed in the collimated beam path and close to the FL. When no voltage is applied to the ETL, the previous configuration/analysis still applies. We have named it as "ETL at 0D" to notice that 0 diopters are produced by the ETL. Then, positive and negative maximum voltages are applied to the ETL which provide the positions "ETL at 6D" and "ETL at -2D", respectively. Positive voltage increases ETL optical power and the point source is shifted away by a distance Δz from the sample. This distance can be easily calculated from

$$\Delta z = \frac{f'^2_{FL}}{f'_{ETL} + f'_{FL}} \qquad (2)$$

being f'$_{FL}$ and f'$_{ETL}$ the focal lengths of FL and ETL, respectively. Equation 2 is not exact since it assumes that both lenses (FL and TL) are thin lenses and are placed on the same axial location. But it provides a first approximation to the theoretical values. In-depth analysis and calibration are included in next section using a resolution test target but a couple of examples are commented here for clarity. Using Eq. 2 and for the positive voltage (6D optical power means that f'$_{ETL}$ = 166.67 mm), Δz results in 0.366 mm approximately. However, the magnification is properly reduced as the new source-sample distance becomes increased: from 26X to 15.4X. Obviously, the contrary happens when a negative voltage (-2D optical power means that f'$_{TL}$ = -500 mm) is applied to the ETL (Δz ~ -0.130 mm and 34.8X, approx.).

Finally, the recorded in-line holograms must be numerically processed to decisively achieve imaging conditions. In that sense, there are different numerical methods for digital reconstruction by solving the diffraction integral [33]. Among them, we have selected to compute the Rayleigh-Sommerfeld diffraction integral using convolution operation since it allows an effective and economical calculation without any approximation [34, 35]. In this way, the diffraction integral can be numerically figured exactly by using three Fourier transformations through the convolution theorem, that is, $RS(x,y;d)=FT^{-1}\{FT\{U(x,y) R(x,y)\}\cdot FT\{h(x,y;d)\}\}$, being $RS(x,y)$ the propagated wave field, $U(x,y)$ the processed hologram resulting from the phase-shifting algorithm, $R(x,y)$ the reference wave, $h(x,y)$ the impulse response of free space propagation (the definition of $h(u,v;d)$ can be found in Ref. [33], page 115, Eq. 3.73), $(x,y)$ the spatial coordinates, *FT* the numerical Fourier transform operation realized with the FFT algorithm, and *d* the propagation distance which can be easily found using different criteria [36, 37]. Since usually the Fourier transformation of the impulse response is defined as $H(u,v;d)=FT\{h(x,y;d)\}$, being the spatial-frequency coordinates $(u,v)$, the calculation of the propagated wave field to an arbitrary distance *d* is simplified according to $RS(x,y;d)=FT^{-1}\{\hat{U}(u,v) H(u,v;d)\}$, where $\hat{U}(u,v)$ is the Fourier transformation of $U(x,y)$.

## 3. Experimental results.

*3.1. Setup calibration*

Experimental demonstration of VZ-DIHM is provided through Fig. 2 where a USAF 1951 resolution test target is used as input object. Figure 2 includes the 3 recorded holograms in (a)-(b)-(c) coming from positive-zero-negative voltages, respectively, as well as the 3 in-focus images in (d)-(e)-(f) after numerical propagation of the 3 in-line holograms. The total FOVs and magnification factors are included in Table 1. Both the FOVs and the magnifications are computed by considering plots along known elements in the test target [dashed white lines at Figs. 2(d) to 2(f)].

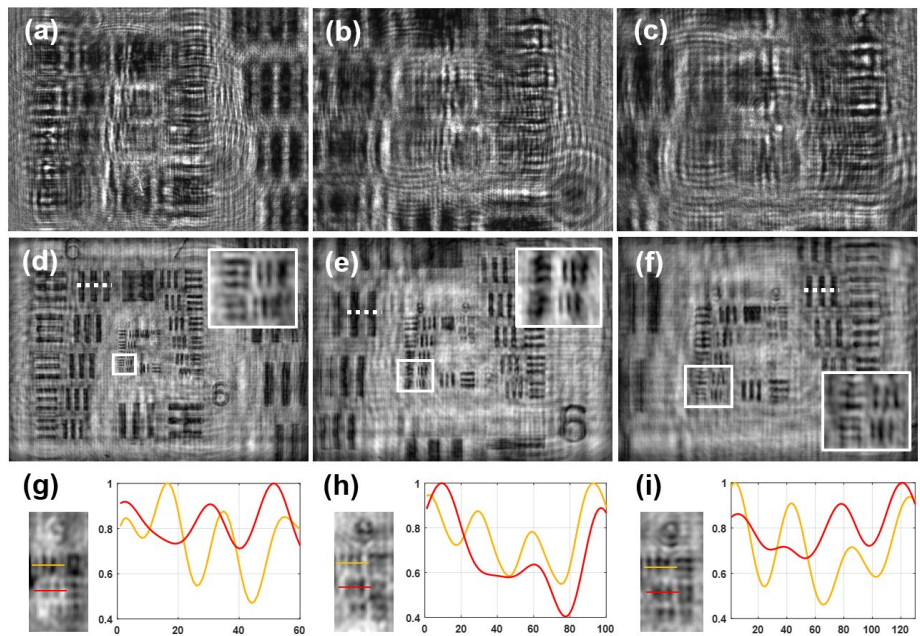

Figure 2. Experimental validation using a resolution test target: (a)-(b)-(c) are the recorded in-line holograms when positive-zero-negative voltages are respectively applied to the ETL; these holograms are numerically focused at (d)-(e)-(f) where the insets are to magnify that the last 2 Elements of the Group 8 are resolved; and finally (g)-(h)-(i) includes the 3 elements of Group 9 where plots along the Elements 1 (orange) and 2 (red) are presented in order to clearly show that the resolution limit is the Element 1 of Group 9 on all cases.

From a resolution limit point of view, we have included a magnified inset corresponding with Elements 5 and 6 of Group 8 (E5/6-G8) [Figs. 2(d) to 2(f)] as well as a magnification of the Group 9 including normalized intensity plots for Elements 1/2 [Figs. 2(g) to 2(i)]. As expected, there is a different resolution limit in horizontal and vertical directions coming from a rectangular sensor area. Vertical bars of E5/6-G8 are clearly resolved and resolution limit comes from E1-G9 for all three magnification cases. Note that we are not changing the NA of the imaging system ($\Delta z$ affects M and FOV); so, in principle it should be the same because the geometrical limit imposed by the sampling criterion is not limiting the resolution in any case. Plots included at Figs. 2(g) to 2(i) clearly show the 3 valleys corresponding with E1-G9 (orange plot) while cannot be visible for E2-G9 (red plot). Thus, E1/G9 defines a resolution limit of 1.95 μm (512 lp/mm) which is in perfect agreement with the theoretical prediction.

Table 1. Experimental FOV and magnification analysis when varying the ETL voltage.

|  | 6D | 0D | -2D |
|---|---|---|---|
| FOV (μm x μm) | 360x270 | 218x163 | 159x119 |
| M factor | 15.6X | 25.9X | 35.4X |

In order to validate the results, Fig. 3(a) plots the magnification theoretical curve (Eq. 1) as a function of the z distance and assuming that d = 12.5 mm. From the plot, the theoretical magnification values for the different ETL voltages are retrieved resulting in 15.4X, 26X and 34.8X for ETL at 6D, 0D and -2D, respectively. Those values are in perfect agreement with the experimentally measured ones (see Table 1). In addition, Fig. 3(b) plots the magnification ratio that the ETL is generating as a function of the z distance. It is defined as the ratio between the magnifications provided by the ETL at maximum/minimum voltages, that is, M(-2D)/M(6D). The curve is interesting to highlight that the shorter the z distance, the higher the magnification ratio. Then, the zoom function without mechanical movement provided by the ETL is applicable in LDHM when the source is in close proximity with the sample (DIHM implementation) while remain useless when the sample is on top of the digital sensor (on-chip microscopy arrangement). In our case, the M ratio equals to 2.26 meaning that the FOV is changed by a factor of $(2.26)^2$.

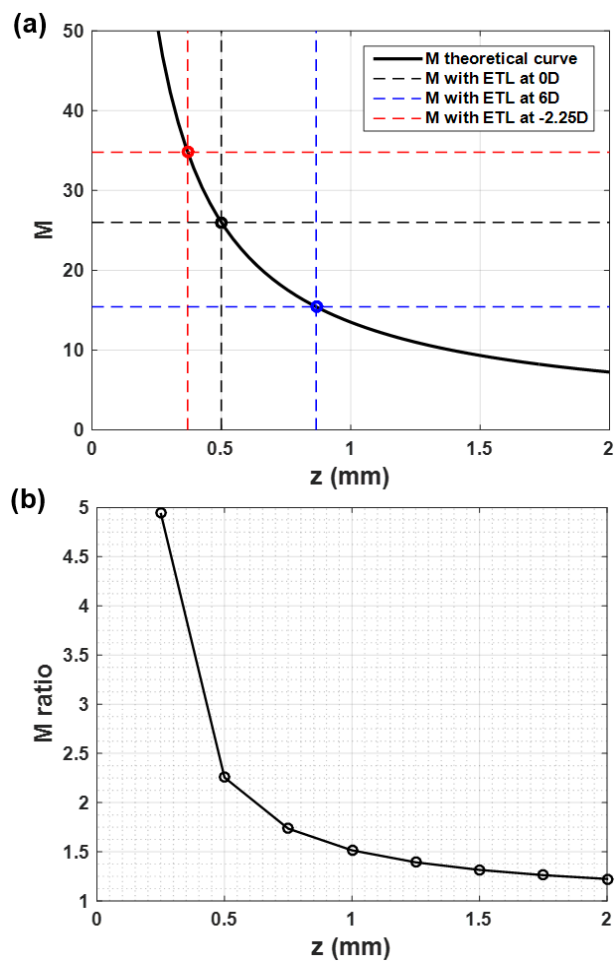

Figure 3. (a) Theoretical plot of the magnification curve and the points caused by the ETL variation; (b) magnification ratio produced by the ETL. Both curves are plotted as a function of the z distance.

*3.2. Biosample validation*

Experimental validation is also provided considering a fixed prostate cancer cell bio-sample. The results are included in Fig. 4 where only numerical reconstructions are included.

Thus, Figs. 4(a) and 4(b) present the intensity images for the ETL at 6D and -2D, respectively. Leaving aside halo effects and twin image disturbances, the zoom function of the ETL permits to optically magnify the central area (marked with a dashed white rectangle) thus allowing high quality image of the cell. In addition, Figs. 4(c)-(d) include the phase retrieved images for the cases where the ETL is set to 0D and removed from the layout, respectively.

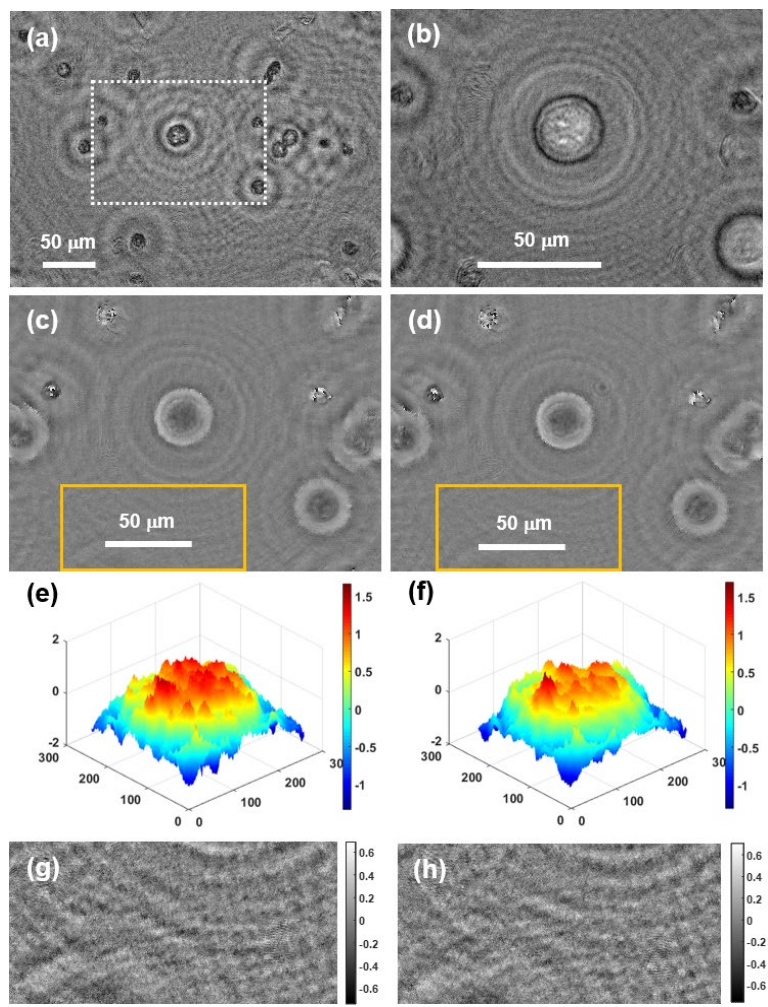

Figure 4. Experimental validation using a prostate cancer cell bio-sample: (a)-(b) are the intensity in-focus images retrieved when the ETL is set to positive (ETL at 6D) and negative (ETL at -2D) voltages, respectively, and where the dashed line rectangle in (a) bounds the roughly same FOV than in (b) for direct comparison; (c)-(d) show the retrieved phase images with the ETL at 0 voltage and removing the ETL from the layout, respectively, for direct comparison of the case with/without ETL in the layout; (e)-(f) include the same

comparative than in (c)-(d) but showing 3D views of the retrieved phase values for the central cell; and finally (g)-(f) are the background phase distributions coming from the orange rectangles in (c)-(d), respectively, used to analyze the STD of the background. Scale bars in (e)-(f)-(g)-(h) represent optical phase in radians.

And 3D map comparative of central cell is presented in Figs. 4(e)-(f) showing good quality phase reconstruction on both cases. This fact is also corroborated by analysing the standard deviation (STD) in the background of the retrieved phase distributions. Thus, Figs. 4(g)-(h) show the phase distribution of a clear background area (marked with orange rectangles in Figs. 4(c)-(d), respectively) from which STD values are computed resulting in 0.0023 rads and 0.00024 rads corresponding with the cases of ETL at 0D and no ETL in the layout, respectively. Those values imply no significant phase variation is induced by the inclusion of the ETL.

Finally and in order to validate the retrieved phase values with the presented variable zoom DIHM platform, a classical DHM platform based on a SMIM interferometric configuration [38] is assembled at the lab and the prostate cancer cell bio-sample is visualized in phase. Figure 5 includes the phase images of the inspected area. Although it is not the same specific area as the one included in Fig. 4, it is the same bio-sample containing cells prepared using the same fixing conditions; so a comparable phase maps can be retrieved and compared. Figure 5 presents different cells having different phase profiles but we have selected 4 of them (marked with coloured squares) providing similar phase values (ranging from -1 to 1.5 rads, approx..) than the ones included in Figs. 4(e) and (f). So, quantitative phase values are in perfect agreement.

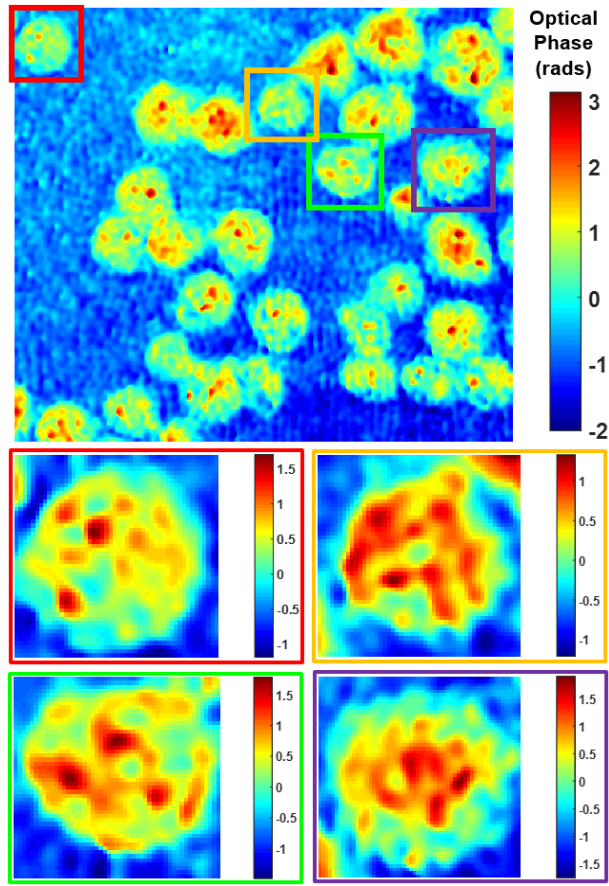

Figure 5. Prostate cancer cells visualized in phase. Experimental results obtained using a lab-made DHM platform based on a SMIM interferometric configuration with a 20X/0.46NA objective lens.

## 4. Discussion and conclusions

Closely related with the presented VZ-DIHM approach, the use of an ETL in LDHM were previously introduced [23] but with a completely different aim: to record the image of an object having a total extension which is larger than that of the image sensor. The authors demonstrated for the first time magnifications lower than 1 in LDHM [23] where an ETL allowed to obtain magnifications ranging from 0.65 to 1.21. Here, we are applying to microscopy field where high magnifications are pursued and the possibility to select a specific sample region to provide higher magnification imaging is an extremely useful capability.

We have selected a system configuration (z = 0.5 mm, d = 12.5 mm) enabling 0.22NA, M = 26X, $\rho$ = 1.85 µm and FOV = 218x163 µm. This configuration is open and can be redefined according to the envisioned specific application. For instance, the NA can be improved by slightly reducing the d distance (for example, from 12.5 to 10 mm). Thus, the magnification will be decreased and the resolution limit and FOV will become improved. When considering these modifications, the system specifications will be really close to the ones provided by a 20X infinity corrected microscope objective. And the ETL will provide continuous zoom variation from 12X to 28X without penalizing the resolution limit because the d distance in unaltered. With the ETL used in the presented implementation, the zoom function ranges from 15X to 35X incoming from a focal tuning range of -2 to +6 diopters. However, there are nowadays new ETLs having higher variation range (for instance, the EL-16-40-TC is capable to change between -10 to +10 diopters), so higher performance will be obtained in a straightforward manner.

Maybe one of the major concerns regarding ETLs is image quality. Presented quantitative phase analysis shows comparable results when including or not the ETL in a DIHM layout. Nevertheless, the comparative included here is at an operating current of 0 mA in the ETL and aberrations can appear when the ETL voltage is increased. In that sense, Deng et al. [31] presented a study about Zernike aberrations introduced by an ETL (the same one that is used in this manuscript) when varying the driven current. They conclude that the ETL does not introduce another phase error except defocus aberration. However, they suggested that the ETL should be mounted horizontally to avoid the gravity effect in the ETL membranes which can cause comma aberration when vertical assembly is considered. Maybe this is the reason causing mismatch in the background STD phase values (1 order of magnitude in difference) since, in our layout, the ETL is mounted vertically. Further studies such as aberrations analysis from the image center towards image periphery and comparison for different driven currents can be performed but this is out of the scope of this paper. Nonetheless, it is reasonable to think that aberrations will increase as the driven current does it, especially when arriving to the ETL limits. In such as cases, the retrieved image quality will depend on the ETL power and will be probably deteriorated towards the borders of the image meaning that the usable FOV will be restricted to the image center. But this is not a critical drawback of the proposed approach since high quality reconstruction at high magnification will still be possible at the expenses of FOV reduction.

The proposed approach belongs to the field of lensless microscopy, more specifically to DIHM where magnification is pursued in the imaging system. DIHM is nowadays an appealed approach with huge potential mainly in the field setting where compact, robust, portable and cost-effective devices are highly demanded in many biological/biomedical applications. Some examples are global healthcare and point-of-care diagnosis [39], continuous monitoring of cellular cultures inside an incubator at a controlled temperature and humidity [40], identification and characterization of 3D particle location within a sample volume [41], portable microscopes with rapid and accurate diagnosis in the field setting [42, 43], quantification of basic cell functions such as cell-substrate adhesion, cell spreading, cell division, cell division orientation and cell death [44], in situ submersible imaging of seafloor environments and organisms [45-47], as background technology for space missions and exobiological studies [48], particle tracking (from intracellular dynamics to the characterization of cell motility and migration) [49-51], and for 3D trajectory analysis in sperm samples [43, 52, 53]. Combination with other well-established techniques in microscopy such as fluorescence [54, 55] or tomography [56] is also possible. Hence, the proposed layout can improve previous developments in a vast range of biomedical applications.

In summary, we have reported on VZ-DIH; as a novel approach in DIHM where the system magnification and FOV is continuously changed without mechanical movement of any component in the layout but rather actively tailoring the wavefront curvature of the illumination beam. This is achieved by inserting an ETL prior to the focusing lens. Thus, an optical zoom function is achieved by activating the ETL allowing a magnification/FOV variation factors of 2.26 and $(2.26)^2$, respectively. Experimental results are provided by using a resolution test target to validate/calibrate the main system parameters and a biological sample to show its potential in biomedical imaging.

**Acknowledgements**

This work has been funded by the Spanish Ministerio de Economía y Competitividad and the Fondo Europeo de Desarrollo Regional (FIS2017-89748-P), and by the National Science Center, Poland (OPUS 13 2017/25/B/ST7/02049), Warsaw University of Technology statutory funds, Polish National Agency for Academic Exchange.

# FIGURE CAPTIONS

Figure 1.

Optical layout for the proposed variable magnification DIHM. CL, condenser lens; ETL, electrically tunable lens; FL, focusing lens.

Figure 2.

Experimental validation using a resolution test target: (a)-(b)-(c) are the recorded in-line holograms when positive-zero-negative voltages are respectively applied to the ETL; these holograms are numerically focused at (d)-(e)-(f) where the insets are to magnify that the last 2 Elements of the Group 8 are resolved; and finally (g)-(h)-(i) includes the 3 elements of Group 9 where plots along the Elements 1 (orange) and 2 (red) are presented in order to clearly show that the resolution limit is the Element 1 of Group 9 on all cases.

Figure 3.

(a) Theoretical plot of the magnification curve and the points caused by the ETL variation; (b) magnification ratio produced by the ETL. Both curves are plotted as a function of the z distance.

Figure 4.

Experimental validation using a prostate cancer cell bio-sample: (a)-(b) are the intensity in-focus images retrieved when the ETL is set to positive (ETL at 6D) and negative (ETL at -2D) voltages, respectively, and where the dashed line rectangle in (a) bounds the roughly same FOV than in (b) for direct comparison; (c)-(d) show the retrieved phase images with the ETL at 0 voltage and removing the ETL from the layout, respectively, for direct comparison of the case with/without ETL in the layout; (e)-(f) include the same comparative than in (c)-

(d) but showing 3D views of the retrieved phase values for the central cell; and finally (g)-(f) are the background phase distributions coming from the orange rectangles in (c)-(d), respectively, used to analyze the STD of the background. Scale bars in (e)-(f)-(g)-(h) represent optical phase in radians.

Figure 5.

Prostate cancer cells visualized in phase. Experimental results obtained using a lab-made DHM platform based on a SMIM interferometric configuration with a 20X/0.46NA objective lens.

Table 1.

Experimental FOV and magnification analysis when varying the ETL voltage.